\documentclass[conference]{IEEEtran}
\IEEEoverridecommandlockouts
\usepackage{cite}
\usepackage{authblk}
\usepackage{lipsum}
\usepackage{amsmath,amssymb,amsfonts}
\usepackage{algorithmic}
\usepackage{graphicx}
\usepackage{textcomp}
\usepackage{xcolor}
\usepackage{threeparttable}
\newcommand{\ineq}[1]{$#1$\normalsize}{}
\def\BibTeX{{\rm B\kern-.05em{\sc i\kern-.025em b}\kern-.08em
    T\kern-.1667em\lower.7ex\hbox{E}\kern-.125emX}}
    
\newcommand{\clarify}[1]{\textcolor{black}{#1}}

\begin{document}
\bstctlcite{IEEEexample:BSTcontrol}

%\title{Respiratory Classification with Deep Learning%\\
%\thanks{Sponsoring funding agency withheld for double blind review process.}
%}
\title{Energy-Efficient  Respiratory Anomaly Detection  in Premature Newborn Infants}

\author{Ankita Paul$^{1}$, Md. Abu Saleh Tajin$^{1}$, Anup Das$^{1}$, William Mongan$^{2}$ and Kapil Dandekar$^{1}$}

\affil{$^{1}$ \quad Department of Electrical and Computer Engineering, Drexel University, Philadelphia, PA 19104 \\
$^{2}$ \quad Department of Mathematics and Computer Science, Ursinus College, Collegeville, PA 19426}

\maketitle

\begin{abstract}
Precise monitoring of respiratory rate in premature newborn infants is essential to initiating medical interventions as required. Wired technologies can be invasive and obtrusive to the patients.
%, and they involve inserting specialized catheters into the infant's tiny veins. 
We propose a Deep Learning enabled wearable monitoring system for premature newborn infants, where \clarify{respiratory cessation is predicted} using signals that are collected wirelessly from a non-invasive wearable Bellypatch put on the infant’s body. We propose a five-stage design pipeline involving data collection and labeling, feature scaling, deep learning model selection with hyper parameter tuning, model training and validation, and model testing and deployment. The machine learning model used is a {1-D Convolutional Neural Network (1DCNN)} architecture with {1} convolution layer, {1} pooling layer, and {3} fully-connected layers, achieving {97.15\%} classification accuracy. To address energy limitations of wearable processing, several quantization techniques are explored and their performance and energy consumption are analyzed for the respiratory classification task. {Results demonstrate a reduction of energy footprints and model storage overhead with a considerable degradation of the classification accuracy, meaning that quantization and other model compression techniques are not the best solution for respiratory classification problem on wearable devices.} 
Hence,to improve classification accuracy, yet reduce the energy consumption we propose a novel Spiking Neural Network (SNN)-based respiratory classification solution, which can be implemented on event-driven neuromorphic hardware platforms. To this end, we propose an approach to convert the analog operations of our baseline trained 1DCNN to their spiking equivalent. We perform a design-space exploration using the parameters of the converted SNN to generate inference solutions having different accuracy and energy footprints. We select a solution that achieves an accuracy of 93.33\% with 18x lower energy compared to the baseline 1DCNN model. Additionally, the proposed SNN solution achieves similar accuracy as the quantized model with a 4x lower energy.
\end{abstract}

\begin{IEEEkeywords}
Wearable, Respiratory Classification, Deep Learning, Spiking Neural Network (SNN)
%wireless communications, testbeds, software defined radio
\end{IEEEkeywords}

\section{Introduction}
\label{sec:intro}
A premature newborn infant is one who is born more than three weeks before the estimated due date. Common health problems of these infants include Apnea of Prematurity (AOP), which is a pause in breathing for 15 to 20 seconds or more~\cite{eichenwald2016apnea} and Neonatal Respiratory Distress Syndrome (NRDS), which is shallow breathing and a sharp pulling in of the chest below and between the ribs with each breath~\cite{clements1998lung}. Precise respiratory monitoring is often necessary to detect AOP and NRDS in premature newborn infants and initiate medical interventions as required~\cite{rocha2018respiratory}. Wired monitoring techniques are invasive and can be obtrusive to the patient.%they involve inserting catheters into the infant's tiny veins.
Therefore, non-invasive respiratory monitoring techniques are recommended by pediatricians to increase comfort of infants and facilitate continuous home monitoring~\cite{antognoli2018assessment}.

\begin{figure}[h!]
\begin{center}  
\includegraphics[width=0.65\columnwidth]{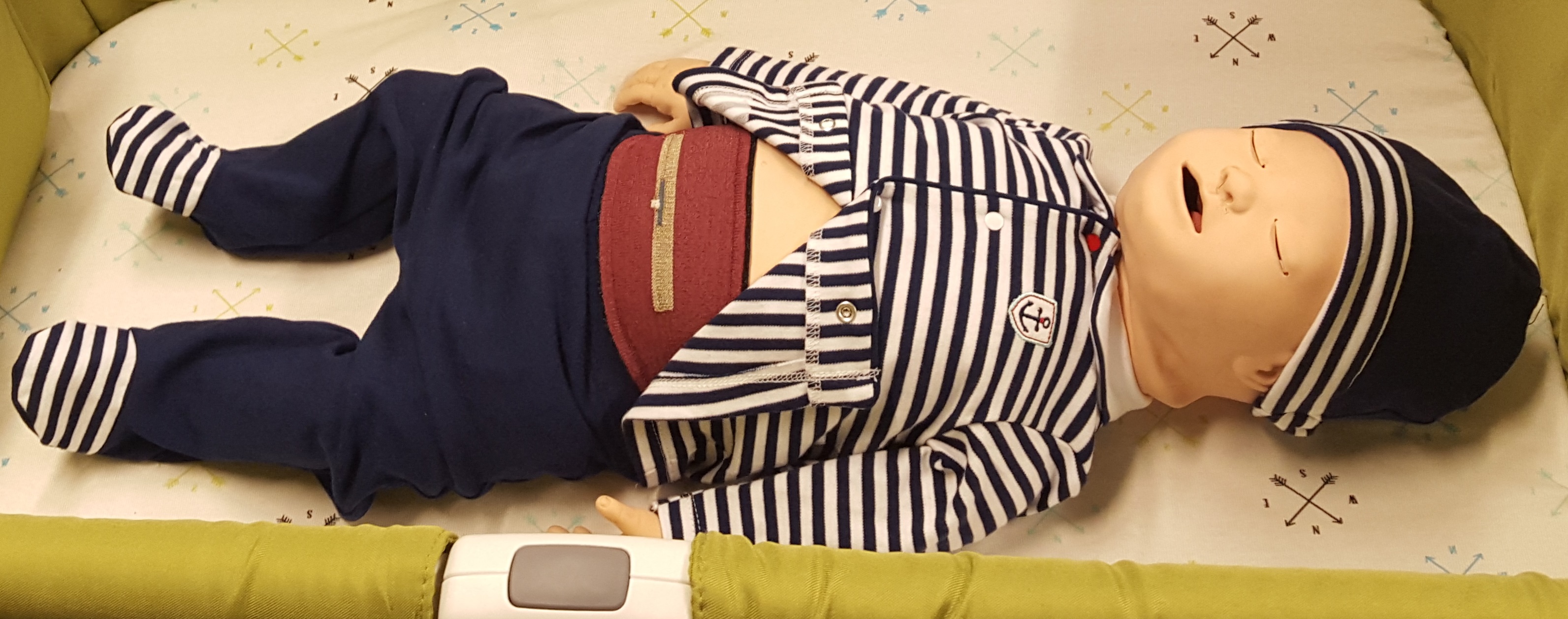} 
\includegraphics[width=0.3\columnwidth]{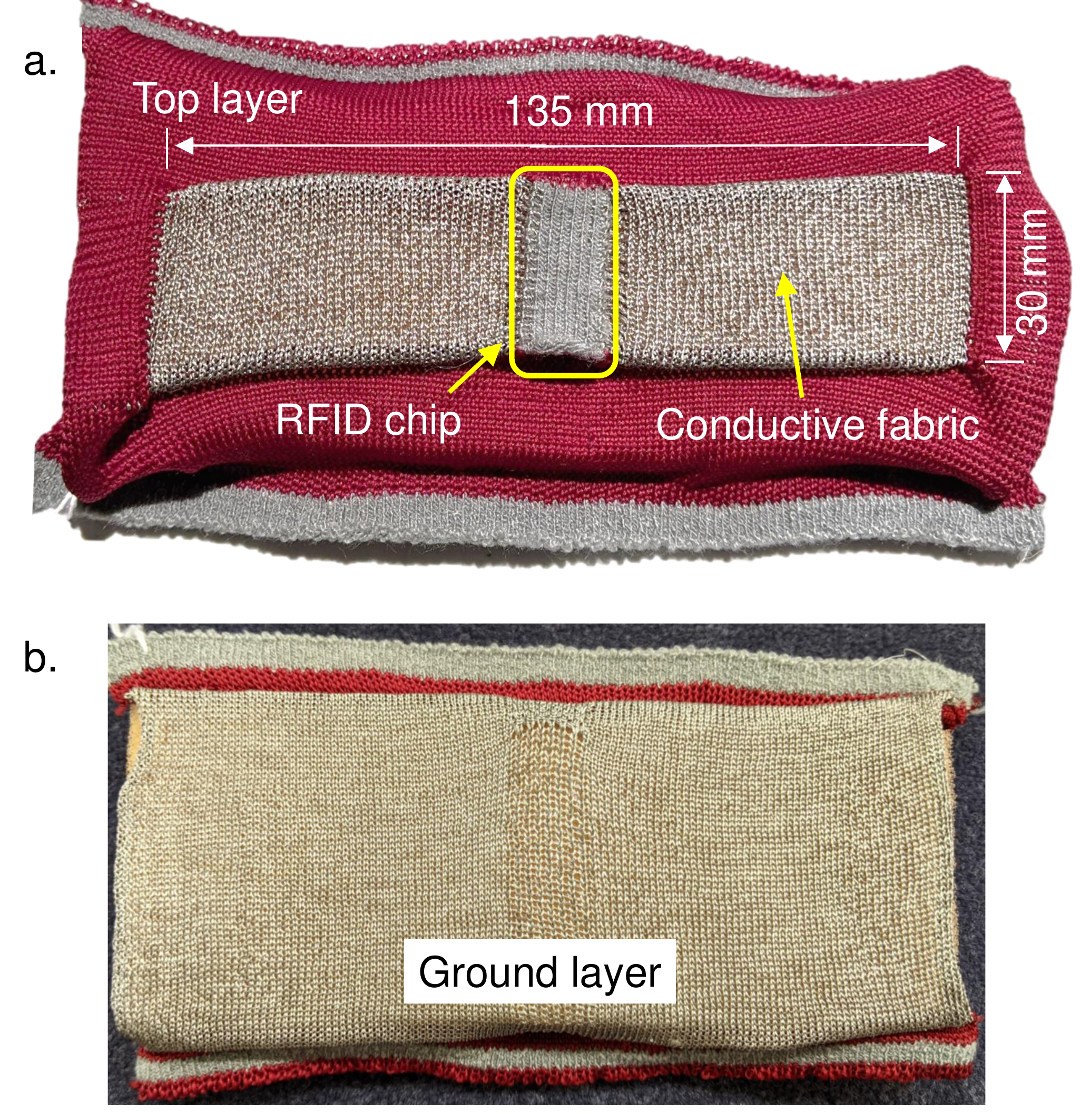} 
\caption{(left) A smart-fabric Bellyband and (right) Bellypatch can be integrated into wearable garments to enable wireless and passive biomedical monitoring in infants.}  
\vspace{-1.5em}
\label{fig:bellyband}
\end{center}  
\end{figure}

We study the use of wearable technologies in respiratory monitoring of infants. To this end, we use the Bellypatch (see Fig.~\ref{fig:bellyband}), a wearable smart garment that utilizes a knitted fabric antenna and passively reflects wireless signals without requiring a battery or wired connection~\cite{7458913,9383785,7501674}.  The Bellypatch fabric stretches and moves as the infant breathes, contracts muscles, and moves about in space; the physical properties of the radio frequency (RF) energy reflected by the antenna change with these movements.  These perturbations in RF reflected properties enable detection and estimation of the infant's state, including respiration rate~\cite{ross2021}, heart rate~\cite{7945586}, movement of the extremities~\cite{8754449}, and detection of diaper moisture~\cite{tajin2020}.  

%\textcolor{red}{From Bill - Kapil and the lab may have a more up-to-date picture with the current-generation antenna; I'm including this as a placeholder for now.}

The Bellypatch operates in the 900 MHz Industrial, Scientific, and Medical (ISM) frequency band using Radio Frequency Identification (RFID) interrogation.  An RFID interrogator emits an interrogation signal multiple times per second (often 30-100 interrogations per second).  In typical RFID technology deployments, only a single interrogation response is needed for inventory purposes; however, many interrogations are sent to overcome collisions among the responding RFID tags and other signal interference or loss.  We exploit this redundant interrogation to sample the state of the antenna each time it is successfully interrogated.  Specifically, we observe changes in the received signal strength indicator (RSSI), phase angle, and successful interrogation rate to estimate respiratory properties of the wearer's state as well as other biosignals.  The use of passive RFID enables a wireless and unobtrusive wearable device that requires no batteries to operate; the interrogation signal itself is sufficient to power the worn RFID tag for each interrogation.  However, path loss, multipath fading, and collision mitigation among the tags in the field require signal denoising and interpolation of the received signal as well as intelligent algorithms for signal processing and estimation.  Because these biomedical estimates require real-time or near-real-time processing, and may take place on low-power embedded or portable systems, it is desirable to utilize techniques that place minimal constraints on power consumption, online training, and processing latency.

To this end, we propose a deep learning-enabled respiratory classification approach. At the core of this approach is a five stage pipeline involving data collection and labeling, feature selection, deep learning model selection with hyper parameter optimization, model training and validation, and model testing and deployment. We use a Laerdal SimBaby programmable infant mannequin (see Figure~\ref{fig:bellyband}) to collect respiratory data using the sensors attached on the Bellypatch.  We use a 1-D Convolutional Neural Network (1DCNN) for classification of the features extracted from the SimBaby sensor data. The 1DCNN model consists of {1} convolution layer, {1} polling layer, and {3} fully-connected layers, achieving {97.15\%} classification accuracy (see Section~\ref{sec:baseline_results}). This is higher than state-of-the-art accuracy obtained using machine learning techniques such as Support Vector Machine (SVM), Logistic Regression (LR), and Random Forest (RF), all of which are proposed for respiratory classification of infants. Hyper parameters of this 1DCNN model are selected using a Grid Search method, which is discussed in Section~\ref{sec:grid_search} and the model parameters are trained using the Backpropagation Algorithm~\cite{LeCun2015deep}.

We show that the energy consumption of our baseline 1DCNN model is considerably higher, which makes it difficult to implement the same on the Bellypatch due to its limited power availability. Therefore, we propose several quantization approaches involving limiting the bit-precision of the model parameters. We show that in order to achieve a significant reduction in energy, the model accuracy can be considerably lower. Therefore, model quantization may not be the best solution to implement respiratory classification on the Bellypatch.

Finally, we propose a novel Spiking Neural Network (SNN)~\cite{maass1997networks} enabled respiratory classification solution, which %converts the analog operations of the trained 1DCNN into their spiking equivalent. The converted SNN model 
can be implemented on event-driven neuromorphic hardware such as TrueNorth~\cite{truenorth}, Loihi~\cite{loihi}, and DYNAPs~\cite{dynapse}. We perform design-space exploration using SNN parameters, obtaining SNN solutions with different accuracy and energy. We select a solution that leads to 93.33\% accuracy with 18x lower energy than the baseline 1DCNN model. This SNN-based solution has similar accuracy as the best performing quantized CNN model with 4x lower energy. 

Overall, the SNN-based approach introduces two additional stages in our design pipeline -- model conversion and SNN parameter tuning, making the overall approach, a seven-stage pipeline. Using this seven stage design pipeline, we show that the accuracy is significantly higher than all prior solutions, with considerably lower energy, making this solution extremely relevant for the battery-less Bellypatch.

The remainder of this paper is organized as follows. Related works on respiratory classification are discussed in Section~\ref{sec:related}. The five-stage design pipeline is described in Section~\ref{sec:methods}. Model quantization techniques are introduced in Section~\ref{sec:quantization}. The SNN approach to respiratory classification is formulated in Section~\ref{sec:snn_classifier}. The proposed approach is evaluated in Section~\ref{sec:results} and the paper is concluded in Section~\ref{sec:conclusion}.

%%\textcolor{red}{Add a paragraph or two describing bellyband, with references and pictures, if possible.}

\section{Related Work}
\label{sec:related}
Recently, machine learning-based respiratory classification techniques have shown a significant promise as enabler for continuous respiratory monitoring of newborn infants. To this end, a Support Vector Machine (SVM)-based classifier is proposed in~\cite{mongan2015statistical} achieving 82\% classification accuracy. 
A Logistic Regression-based classifier is proposed in~\cite{acharya2018ensemble} achieving classification accuracy of 87.4\%.
An Ensemble Learning with Kalman filtering is proposed in~\cite{acharya2018ensemble} achieving 91.8\% classification accuracy. All these techniques are proposed for respiratory classification using pulseoximeter data collected from infants, making these approaches relevant state-of-the-art for our work. In Section~\ref{sec:results}, we compare our approach to these state-of-the-art approaches and show that the proposed approach is considerably better in terms of both classification accuracy and energy.
Thermal imaging is also proposed recently for repiratory classicication from infant~\cite{navaneeth2020deep}. The authors reported a precision and recall score of 0.92. We achieve a score of 0.98. Respiratory classification using acoustic sensors are proposed in~\cite{basu2020respiratory}. An accuracy of 95.7\% is reported. We achieved an accuracy of 97.15\%.

Beyond respiratory classification, deep learning enabled techniques have been used extensively for health informatics~\cite{ravi2016deep}. For instance, sleep apnea classification is proposed using deep convolutional neural network (CNN) and long short term memory (LSTM) in~\cite{van2018automated}, achieving an accuracy of 77.2\%. A deep learning approach using InceptionV3 CNN model is proposed in~\cite{henaff2015deep} to detect Alzheimer disease using brain images. Authors reported an area-under-curve (AUC) score of 0.98. CNN models are used in~\cite{bejnordi2017diagnostic} to detect metastatic breast cancer in women. Authors reported an AUC score of 0.994. A CNN-based Arrhythmia classification is proposed in~\cite{sannino2018deep}. Authors reported significant improvement in classification accuracy over state-of-the-art.

Finally, many recent SNN-based techniques have shown comparable and in some case, higher accuracy than their deep learning counterparts with significantly lower energy. An unsupervised SNN-based heartrate estimation is proposed in~\cite{HeartEstmNN}. Authors reported 1.2\% mean average percent error (MAPE) with 35x reduction in energy. A spiking CNN architecture is proposed in~\cite{jolpe18} to classify heart-beats in human. Authors reported 90\% reduction in energy with only 1\% lower accuracy than a conventional CNN. SNN-based epileptic seizure detection is proposed in~\cite{masquelier2020epileptic}. Authors reported an accuracy of 97.6\% with a considerable reduction in energy.

To the best of our knowledge, this is the first work that uses SNN for respiratory classification in infants and show that SNNs can achieve high accuracy (93.33\% in our evaluation) with a considerable reduction in energy (18x lower energy).

%\textcolor{red}{Add more here.}

\section{Design Pipeline}
\label{sec:methods}
\begin{figure}[h!]
    \centering
	%\begin{center}
	    %\vspace{-40pt}
		\includegraphics[width=0.99\columnwidth]{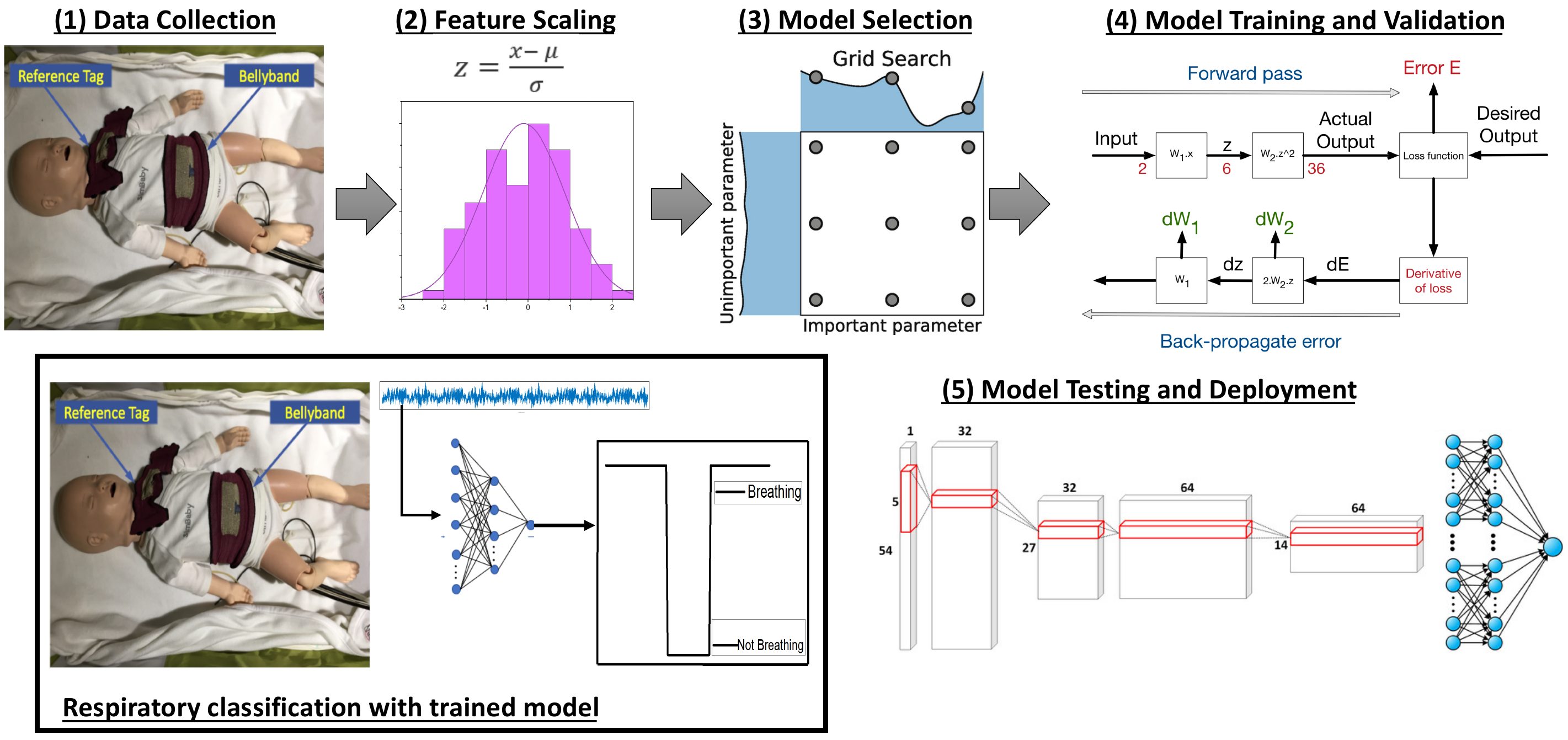}
		%\vspace{-20pt}
	%\end{center}
	\caption{Design pipeline for respiratory classification using Deep Learning.}% related to this proposal.}
	\label{fig:pipeline}
\end{figure}

Figure~\ref{fig:pipeline} shows a high-level overview of the proposed respiratory classification approach using Deep Learning techniques. The design pipeline comprises of five stages -- 1) data collection, 2) feature selection, 3) deep learning model selection, 4) model training and validation, and 5) model testing and deployment. These stages are clearly indicated in the figure. Once a trained model is obtained using this approach, the model is used for respiratory classification of streaming \clarify{pulseoximeter} data collected from the sensors on the Bellypatch. This is shown at the bottom-left corner of Figure~\ref{fig:pipeline}.

In the following, we describe the design pipeline stages.

\subsection{Data Collection and Labeling}
\subsubsection{Data Collection}
In this work, we use a Laerdal SimBaby programmable infant mannequin. The mannequin was programmed to breathe at a rate of 31 breaths per minute for variable time intervals, then to stop breathing for 30 seconds, 45 seconds, and 60 seconds, alternating between these states for a period of one hour. An RFID investigator (Impinj Speedway R420) was used to poll the Bellypatch wearable RFID tag and the antenna with a 900 MHz band RFID signal coming from the SimBaby. The RFID interrogator was also used to measure properties of the backscattered signal reflected from the RFID tag. The interrogator was positioned 1 feet from the mannequin, oriented above, astride, and at the feet. Interrogations were performed with a frequency of 90 Hz.
RFID properties considered for model features include the Received Signal Strength Indicator (RSSI), phase angle, Doppler shift, interrogation frequency, and timestamp.

Each of these properties is affected in-band by the frequency of the original signal emitted by the interrogator. Under United States Federal Communications Commission (FCC) regulations, RFID interrogations must iterate (or channel hop) over 50 frequency channels in the 900 MHz band. In addition to perturbing the raw measurement observations at the interrogator, channel hopping poses challenges in computing higher order features from changes in the observed phase, because these features depend on observing changes in successive values of the phase under the assumption that they were observed from the same interrogation frequency. As a result, the observed Doppler shift is used to identify fine movements of the RFID tag, either in space or because of a strain force applied to the surrounding knit antenna. 

The received signal strength from an interrogation is influenced by several factors as defined by the Radar Cross Section (RCS) formula in Equation~\ref{eq:received_signal}~\cite{su2010investigation}. Specifically, the RCS relates changes in received signal power ($P_{Rx}$) to the interrogation power ($P_{Tx}$), the reader and tag gains ($G_{reader}$ and $G_{tag}$, respectively), the return loss (R), and the interrogation wavelength $\lambda$~\cite{acharya2018ensemble}.

\begin{equation}
    \label{eq:received_signal}
    \footnotesize P_{Rx} = P_{Tx} \cdot G_{reader}^2\cdot G_{tag}^2 \cdot R\cdot \left(\frac{\lambda}{4}\cdot \pi \cdot r \right)^4
\end{equation}

Some of these terms can be controlled by the interrogator configuration: for example, the interrogation transmitter power and antenna reader gain at the interrogator; however, the interrogation frequency changes due to channel hopping, and the receiving antenna gain ($G_{tag}$) changes as the wearer stretches the antenna or moves about in space.  Thus, the observed RSSI alone confounds several artifacts about the state of the transmission along with the state of the wearer.  As a result, a higher order feature $\zeta$ is computed from the RSSI measure by accounting for the interrogation frequency.

We manipulate the RCS equation to arrange those terms related to wearer state on one side, and set them equal to those terms related to the interrogator configuration, as shown in Equation~\ref{eq:gain_change}. Thus, we observe that the changes in the gain of the tag $G_{tag}$ (resulting from movement or a strain force on the antenna), the distance r between the interrogator and the tag (resulting from movement), and return loss $R$ (resulting from movement, strain force, fading or multipath interference) are proportional to the interrogation wavelength lambda and observed $RSSI$ measure $P_{Rx}$, along with the interrogation power $P_{Tx}$ and the reader gain $G_{reader}$, which are held constant at the interrogator and interrogating antenna~\cite{7458913}.
We define $\zeta$ as the terms of RCS model indicative of the changes in the tag, such as movement, strain force, or interference.

\begin{equation}
    \label{eq:gain_change}
    \zeta = G_{tag}^{-2}\cdot r^4\cdot R^{-1} = P_{Tx}\cdot G_{reader}^2\cdot P_{Rx}^{-1}\cdot\left(\frac{\lambda}{4\pi}\right)^4,
\end{equation}

We remove a residual term $\delta = -10 log_{10}\frac{f^4}{(f-(0.5*10^6 MHz))^4} \approx -0.00941$ to compensate for a sawtooth artifact resulting from quantization of the observed RSSI as the interrogation frequency changes among 50 discrete channels per FCC regulations in the United States.%, as the frequency changes by 500 kHz on each change in channel. This constant linear oscillatory pattern is mitigated by computing $\zeta = \zeta - \delta\cdot(50-\omega)$, where $\omega$ is the channel number in $[0, 50)$ that denotes the frequency iteration with respect to the frequency $f$. The channel number is computed using the interrogation frequency as \textcolor{black}{$\omega = \frac{f - 902.25 MHz}{500 kHz}$} to determine which of the 500 kHz offsets from the start of the 900 MHz RFID band is in use.

In summary, we chose the following features for consideration during wireless respiratory state classification.
\begin{itemize}
    \item \textbf{Feature 1:} Reflected signal strength as measured at the interrogator, $\zeta$ (RSSI)
    \item \textbf{Feature 2:} Tag velocity as a function of the observed Doppler shift at the interrogator ($P_{Rx-deoscillated}$)
\end{itemize}
We normalized RFID signal strength (RSSI) data by frequency and calculated the tag velocity to utilize the signal for respiratory analysis. The resulting time-series data was filtered, and signal processed to determine the mean power spectral density, derived from the amplitude of the oscillatory behavior observed in the signal during short time windows.

\begin{figure}[h!]
    \centering
	%\begin{center}
	    %\vspace{-40pt}
		\includegraphics[width=0.99\columnwidth]{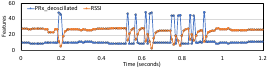}
		%\vspace{-20pt}
	%\end{center}
	\caption{Variation of the two features for 1.2 seconds.}% related to this proposal.}
	\label{fig:features}
\end{figure}

Figure~\ref{fig:features} illustrates the two features ($P_{Rx-deoscillated}$ and RSSI) over a time window  of 1.2 seconds.

\subsubsection{Data Labeling} In case of supervised learning of human-activity recognition from sensor data it is necessary to appropriately label the output. The dataset contained approximately 1685 samples per minute, obtained for 60 minutes resulting in approximately 1 sample generated every 0.03 seconds. This indicated that the RFID interrogation frequency was approximately 28 Hz per RFID tag. For each sample we have two features --- RSSI and $P_{Rx-deoscillated}$ . 

\begin{figure}[h!]
    \centering
	%\begin{center}
	    %\vspace{-40pt}
		\includegraphics[width=0.99\columnwidth]{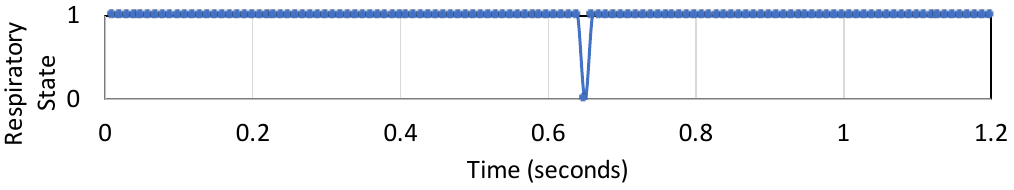}
		%\vspace{-20pt}
	%\end{center}
	\caption{Respiratory state corresponding to features shown in Fig.~\ref{fig:features}.}% related to this proposal.}
	\label{fig:states}
\end{figure}

The observations were broken into time windows of 1 second with no overlap. Hence, each time window contained approximately 28 samples, with two sets of observations from the 2 features. We manually labeled the data collected from the two features as `1' when the SimBaby is in breathing state and `0' when it is in a non-breathing state. Over a period of one hour, we collect and label the dataset to form train and test samples representing binary respiratory state (1: Breathing, 0: Non- Breathing state). Figure~\ref{fig:states} shows the respiratory state corresponding to the features of Figure~\ref{fig:features} for 1.2 seconds.

Since there are only two features and two output classes, the problem we aim to solve is a \textit{bivariate time series binary classification} one.

\subsection{Feature Scaling}
From the time series features extracted from each time window we apply feature engineering to make the input vectors suitable for the classifier. For multivariate data, it is necessary to transform features with different scales to have uniform distribution, to ensure optimal performance of the classifiers. We first cleaned our feature set by filtering the missing values (NaN). The data with the features and the labels were loaded from two csv files and then  we split the dataset 3:1 to form the training set and the testing set. After splitting the dataset, we scale the features before we fit it into our classifier, which is a one dimensional convolutional neural network (1DCNN).

The two features in our dataset were scaled to a standard range, the distribution of the values was rescaled, so the mean was 0 and the standard deviation was 1. 
The method involved determining the distribution of each feature and subtracting the mean from each feature. Then we divide the values (after the mean has already been subtracted) of each feature by its standard deviation.

The standard score ($Z$) of a sample is given by Equation~\ref{eq:standard_score}.
\begin{equation}
    \label{eq:standard_score}
    Z = \frac{x-\mu}{\sigma},
\end{equation}
where $x$ is the sample value, and $\mu$  and $\sigma$ are the mean and standard deviation of all the samples, respectively.
Feature standardization transforms the raw values into the standard scale that helps the model to extract salient signal information from the observations.
 After rescaling the variables, we reshape the data according to dimension expected by the convolution layer of the 1DCNN model. 
 
 \subsection{Deep Learning Model Selection}
 A Convolution Neural Network (CNN) is a class of deep learning that uses a linear operation called convolution in at least one of their layers. Equation~\ref{eq:convintegral} represents a convolution operation, where $x$ is the input and $w$ represents the kernel, which stores parameters for the model. The output $s$ is called the feature map of the convolution layer.
 
 \begin{equation}
    \label{eq:convintegral}
        \footnotesize s(t) = \int x(a)w(t - a)da
\end{equation}

 In CNN, the first layer is a convolution layer that accepts a tensor as an input with dimensions based on the size of the data \cite{Goodfellow2016Deep}. The second layer, or the first hidden layer, is formed by applying a kernel or filter that is a smaller matrix of weights over a receptive field, which is a small subspace of the inputs. Kernels apply an inner product on the receptive field, effectively compressing the size of the input space \cite{LeCun2015deep}. As the kernel strides across the input space, the first hidden layer is computed based on the weights of the filter. As a result, the first hidden layer is a feature map formed from the kernel applied on the input space. While the dimension of the kernel may be much smaller in size compared to the initial inputs of the convolution layer, the kernel must have the same depth of the input space. The inputs and convolution layers are often followed by rounds of activation, normalization, and pooling layers \cite{LeCun2015deep}. The precise number and combination of these layers are specific to the problem at hand. %The last layer, however, is a fully connected layer where the final outputs or categorizations are determined based on how different features fall in line with the specific classes under study. 

\begin{figure}[h!]
    \centering
	%\begin{center}
	    %\vspace{-40pt}
		\includegraphics[width=0.99\columnwidth]{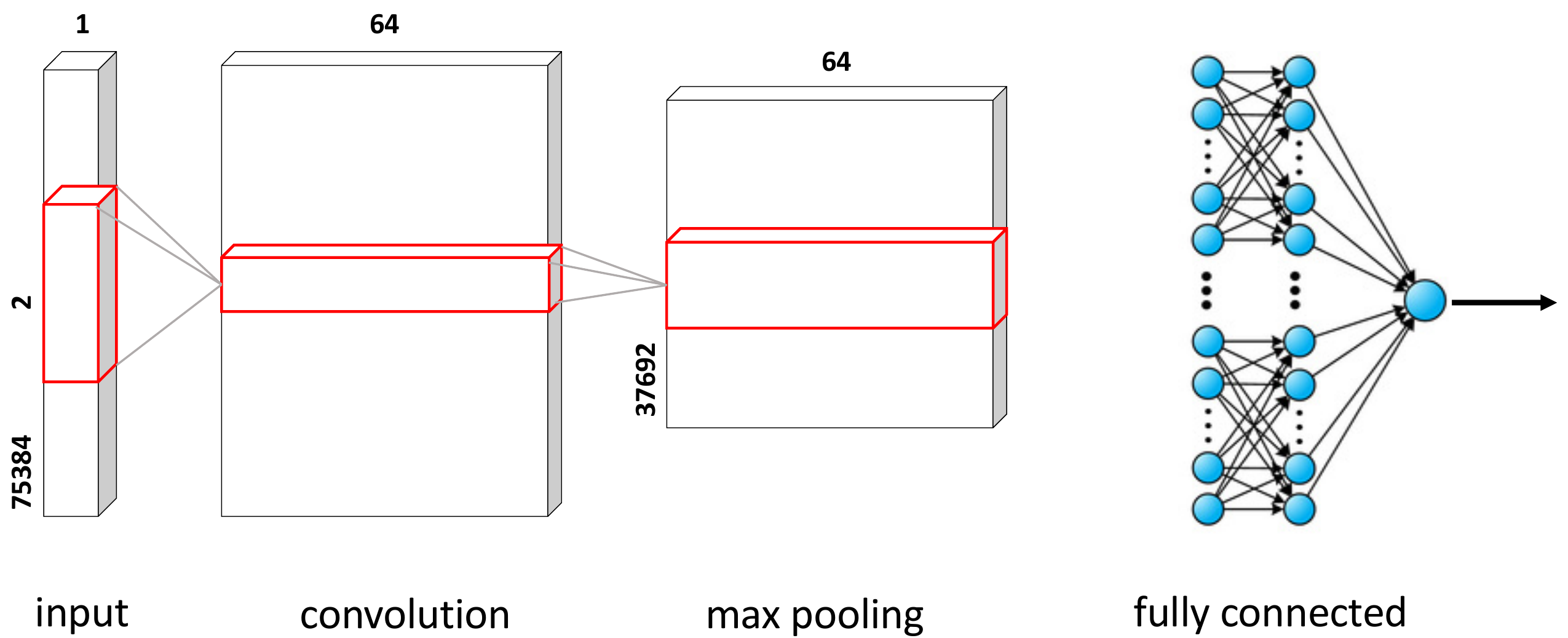}
		%\vspace{-20pt}
	%\end{center}
	\caption{Our proposed 1DCNN architecture.}% related to this proposal.}
	\label{fig:1dcnn}
\end{figure}

For the proposed respiratory classification problem, our CNN model consists of one convolution layer, which is activated by a rectified linear unit (ReLU). The filter size is set to 64 and the kernel size to 1. This layer is followed by a one dimensional Max Pooling layer with a pool size and stride length of 1, each. The next layer is a Flattenning layer followed by a Dropout layer. The Dropout layer randomly sets input neurons to 0 with a rate of 0.01 at each step during training time. This is done to prevent overfitting. The dropout layer is followed by two fully connected hidden layers. The first hidden layer consists of 200 neurons with ReLU activation, the second hidden layer contains 100 neurons with a Softmax activation function. Overall, the proposed CNN model uses one-dimensional convolutions and therefore, this model is referred to as \textit{1DCNN}. Figure~\ref{fig:1dcnn} shows the proposed 1DCNN architecture along with the dimension of each layer.

\subsection{Hyperparameter Optimization}\label{sec:grid_search}
In a CNN, the parameters in each layer whose values control the learning procedure are called hyperparameters. Grid search is a hyperparameter tuning technique that can build a model for every new combination of hyperparameters that are specified in the search space and evaluates each model for that combination. 
A machine learning algorithm $A_\theta$ can estimate a model $M_\theta \in M$ that minimizes a loss function $\mathcal{L}$ with its model regularization term $\mathcal{R}$ given by
\begin{equation}
    \label{eq:model_regularization}
    \footnotesize A_\theta(X_\text{train}) = \underset{M_\theta \in M}{\texttt{argmin}}\mathcal{L}(M_\theta,X_\text{train}) + \mathcal{R}(M_\theta,\theta),
\end{equation}
where $X_\text{train}$ is the training dataset, $M$  is the set of all models, $\theta\in \Theta$ is the chosen hyperparameter configuration, and $\Theta = \{\Theta_1,\Theta_2,\cdots,\Theta_p\}$ is the p-dimensional hyperparameter space of the algorithm. The optimal hyperparameter configuration $\theta^*$ is calculated using the validation set as
\begin{equation}
    \label{eq:optimal_parameter}
    \footnotesize \theta^* = \underset{\theta \in \Theta}{\texttt{argmin}} \mathcal{L} (A_\theta (X_\text{train},X_\text{validation}) = f_D(\theta),
\end{equation}
where $X_\text{validation}$ is the validation set and $f_D$ is the misclassification rate.

\begin{figure}[h!]
    \centering
	%\begin{center}
	    %\vspace{-40pt}
		\includegraphics[width=0.99\columnwidth]{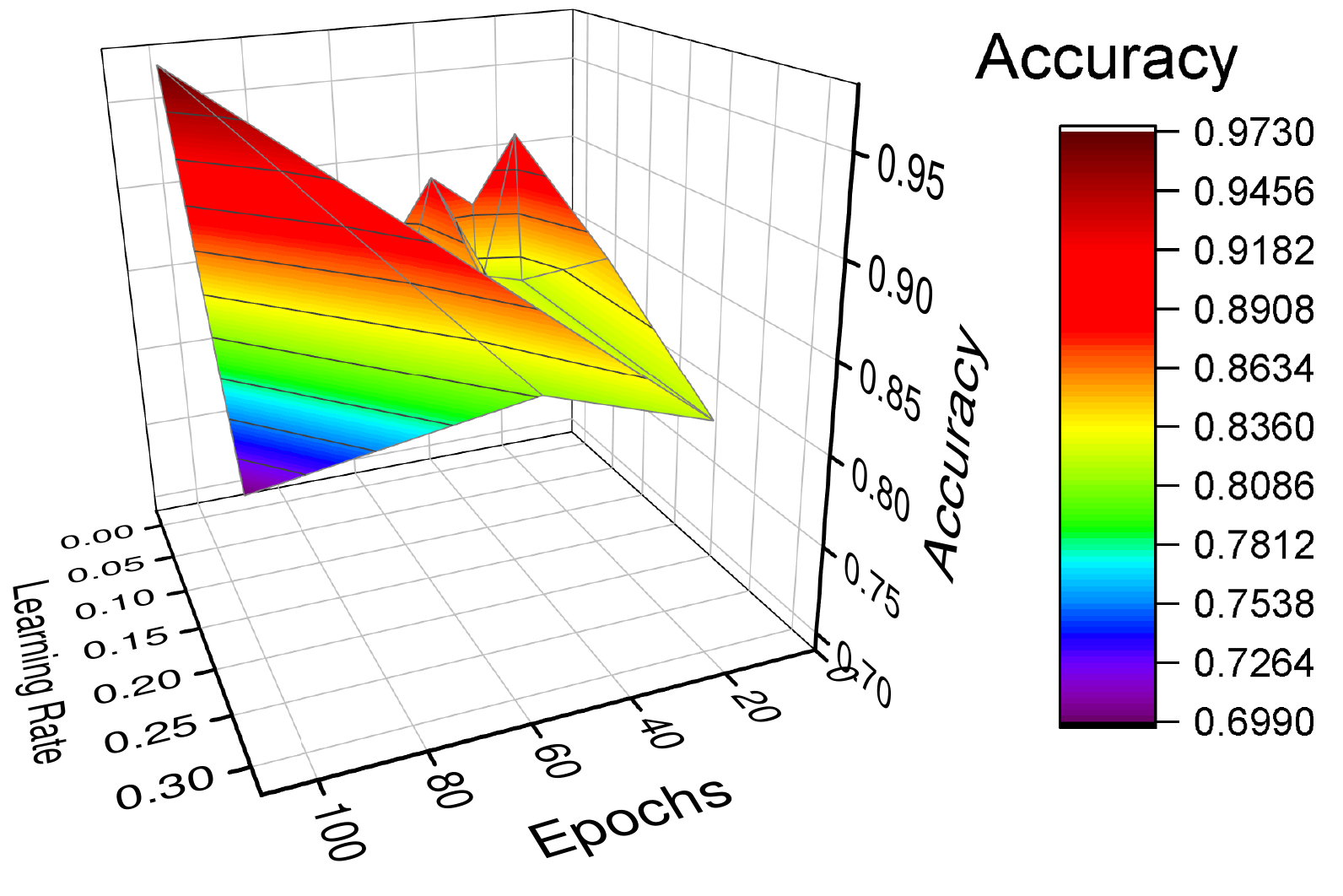}
		%\vspace{-20pt}
	%\end{center}
	\caption{Hyperparameter selection.}% related to this proposal.}
	\label{fig:hyper_parameters}
\end{figure}

The Grid search exhaustively searches through a grid of manually specified set of parameter values provided in a search space to find the accuracy obtained with each combination. We tuned the model based on the number of epochs (ranging from 10 to 100) and the learning rate for the Adam optimizer (0.001, 0.01, 0.1, 0.002, 0.02, 0.2, 0.003, 0.03, 0.3). We evaluated the accuracy as a performance metric for the different combinations.   
Figure~\ref{fig:hyper_parameters} shows the hyperparameter selection using Grid search, where the accuracy is reported for different combinations of epochs and learning rate~\cite{bergstra2012random}.

Table~\ref{tab:hyper_parameters} summarizes the hyperparameters of the 1DCNN.
\begin{table}[h!]
    \caption{Summary of hyperparameters for the proposed 1DCNN model.}
	\label{tab:hyper_parameters}
	\vspace{-5pt}
	\centering
	{\fontsize{10}{12}\selectfont
		\begin{tabular}{lp{5cm}}
			\hline
			Learning rate & 0.001\\
			\hline
			Batch size & 5\\
			\hline
			Optimizer & Adam\\
			\hline
			Data shuffle & per epoch\\
			\hline
			Maximum epochs & 100\\
			\hline
	\end{tabular}}
\end{table}

%The values set for the  other parameters are as follows: learning rate = 0.001, batch size = 5, optimizer = ‘Adam’ [6] data shuffle = every-epoch, and max epochs = 100. The hyperparameters were optimized based on the training/validation performance using Grid Search, explained in the next section. 

\subsection{Model Training and Validation}
We trained our 1DCNN model with 75,834 samples and used repeated k-fold cross-validation with 10 splits to validate our model performance. To improve the estimated performance, we repeated the cross-validation procedure multiple times and reported the mean result across all folds from all runs. This mean accuracy reported is expected to be a more accurate estimate of the true unknown underlying mean performance of the model on the dataset instead of a single run of 
k-fold cross-validation ensuring less statistical noise. We also compute the standard error that provides an estimate of a given sample size of the amount of error that is expected from the sample mean to the underlying and unknown population mean.
The standard error is calculated as
\begin{equation}
    \label{eq:standard_error}
    \sigma_\text{error} = \frac{\sigma}{\sqrt{n}},
\end{equation}
where $\sigma$ is the sample's standard deviation and $n$ is the number of repeats.
We obtained a validation classification accuracy of 87.78\% with a standard error of 0.002 (see Section~\ref{sec:results} for detailed evaluation).
We defined Early Stopping as a regularization technique at the very beginning of declaring the model architecture. At end of every epoch the training loop monitors whether the validation loss is no longer decreasing and once it is found no longer decreasing, the training is terminated. We enabled the patience parameter equal to 5 to terminate the training after epochs of no validation loss decrease.  This is another measure to prevent the model from overfitting during training, alongside the addition of a Dropout layer. Without Early Stopping the training would terminate only after the maximum number of epochs is reached.

\subsection{Model Testing and Deployment}
We deployed the trained model to test the performance on an unseen test set generated from the simbaby to classify the respiratory states. The model was tested on 25,279 samples and achieved an accuracy of 97.15\%, F1 Score of 0.98, AUC score of 0.98, sensitivity score of 0.96 and specificity score of 0.99. These performance metrics are defined in Section~\ref{sec:baseline_results}.

\section{Model Quantization}
\label{sec:quantization}
CNN models consume a considerable amount of energy due to their high computational complexity. 
The high energy consumption is due to the large model size involving several thousand parameters.
It is therefore challenging to deploy the inference, i.e., a trained CNN model on battery-powered mobile devices, such as smartphones and wearable gadgets due to their limited energy budget. To address this high energy overhead, energy-efficient solutions have been proposed to reduce the model size and computational complexity. Some common approaches include pruning of network weights~\cite{liu2020layerwise} and low bit precision networks~\cite{choukroun2019low}. We focus on the latter techniques. Specifically, we implement both bit precision weights and activations to reduce model sizes and computational power requirements.
To perform the training of a CNN with low-precision weights and activations we use the following
quantization function to achieve a k-bit quantization~\cite{coelho2021automatic}.
\begin{equation}
    \label{eq:quantization}
    Z_q = Q(Z_r) = \frac{1}{2^k - 1}\texttt{round}\left((2^k - 1)Z_r\right),
\end{equation}
where \ineq{Z_r \in[0,1]} is the full precision value and \ineq{Z_q\in [0,1]} is the quantized value obtained using the k-bit quantization.

The quantization of weights is given by 
\begin{equation}
    \label{eq:quantization_weights}
    w_q = Q\left(\frac{\texttt{tanh}(w_r)}{2\cdot\texttt{max}\left(|\texttt{tanh}(w_r)|\right)}+\frac{1}{2}\right),
\end{equation}
where \ineq{w_r} is the original weight using full precision and \ineq{w_q} is the quantized value using k-bit quatization.

The quantization of activations is given by 
\begin{equation}
    \label{eq:quantization_activation}
    x_q = Q\left(f(x)\right),
\end{equation}
where \ineq{f(x) = \texttt{clip}(x_r ,0,1)} is the clip function bounding the activation function between 0 and 1.

In this paper we apply both bit precision techniques for both weights and activations using quantization with the QKeras library, which is a quantization extension to the Keras~\cite{keras}.
It enables a drop-in replacement of layers that are responsible for creating parameters and activation layers like the Conv 1D, Dense layers. It facilitates arithmetic calculations by creating a deeply quantized version of a Keras model. We tag the variables, weights and biases created by the Keras implementation of the model and the output of arithmetic layers, by quantized functions. Quantized functions are specified as layer parameters and then passed as a cumulative quantization and activation function, QActivation. The quantized bits quantizer used above performs mantissa quantization using the following equation.
\begin{eqnarray}
    \label{eq:mantissa_quantization}
    &&\text{mantissa quantization} =\\ &&2^{b-k+1}\cdot\texttt{clip}\left(\texttt{round}(x_r\cdot 2^{k-b-1}),-2^{k-1},2^{k-1}-1\right)\nonumber
\end{eqnarray}
where \ineq{x} is the input given to the model, \ineq{k} is the number of bits for quantization, and \ineq{b} specifies how many bits of the bits are to the left of the decimal point. 

We conduct our experiment to perform quantization of our Conv1D model using 2 bits, 4 bits, 8 bits, 16 bits, 32 bits and 64 bits. We observe the performance accuracy increase with the increase of the quantization bits, with 2 bits achieving an 88.93\% compared to using all the 64 bits achieving 97.15\% (see the detailed results in Section~\ref{sec:quantization_results}).

The QTools functionality is used to estimate the model energy consumption for the different bit-wise quantization implementations. It estimates a layer wise energy consumption for memory access and MAC operations in a quantized model derived from QKeras. This is helpful when comparing power consumption of more than one models running on the same device. The model size is calculated as the number of model parameters multiplied by the number of bits used in each scenario. We observe that when we increase in the number of bits, the model size increases, the accuracy increases but also does the consumption of energy (pJ). This homogeneous replacement technique of Keras layers, with heterogeneous per-layer, per-parameter type precision, chosen from a wide range of quantizers, enabled quantization-aware training and energy-aware implementation to maximize the model performance given a situation  of resource constraints, like detection of respiratory cessation on premature infants in critical care conditions, which is crucial for high-performance inference on wearables.

\section{SNN-Based Respiratory Classification}\label{sec:snn_classifier}
\begin{figure}[h!]
    \centering
	%\begin{center}
	    %\vspace{-40pt}
		\includegraphics[width=0.99\columnwidth]{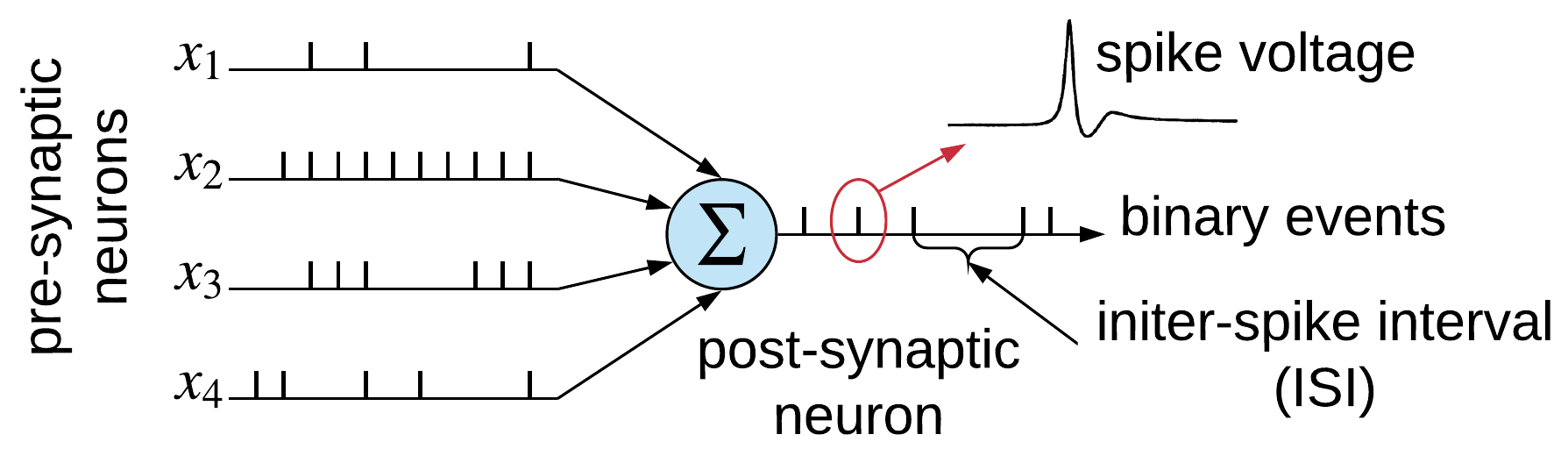}
		%\vspace{-20pt}
	%\end{center}
	\caption{Integration of spike trains at the post-synaptic neuron from four pre-synaptic neurons in a Spiking Neural Network (SNN). Each spike is a voltage waveform of ms time duration.}% related to this proposal.}
	\label{fig:snn}
\end{figure}

Spiking Neural Networks (SNNs), also known as the third generation of neural networks, are interconnection of integrate-and-fire neurons that emulate the working principle of a mammalian brain~\cite{maass1997networks}.
SNNs enable powerful computations due to their spatio-temporal information encoding capabilities.
In an SNN, spikes (i.e., current) injected from pre-synaptic neurons raises the membrane voltage of a post-synaptic neuron. When the membrane voltage crosses a threshold, the post-synaptic neuron emits a spike that propagates to other neurons. Figure~\ref{fig:snn} shows the integration of spike train from four pre-synaptic neurons connected to a post-synaptic neuron via synapses.

%Due to their event-driven activation, SNNs are energy-efficient in implementing 
SNNs can implement many machine learning approaches such as supervised, unsupervised, reinforcement, few-shot, and lifelong learning. 
Due to their event-driven activation, SNNs are particularly useful in energy-constrained platforms such as wearable and embedded system. Recent works demonstrate significant reduction in memory footprint and energy consumption in SNN-based heart-rate estimation~\cite{HeartEstmNN}, heartbeat classification~\cite{das2018heartbeat,jolpe18}, speech recognition~\cite{dong2018unsupervised}, and image processing~\cite{sengupta2019going}. 

\begin{figure}[h!]
    \centering
	%\begin{center}
	    %\vspace{-40pt}
		\includegraphics[width=0.99\columnwidth]{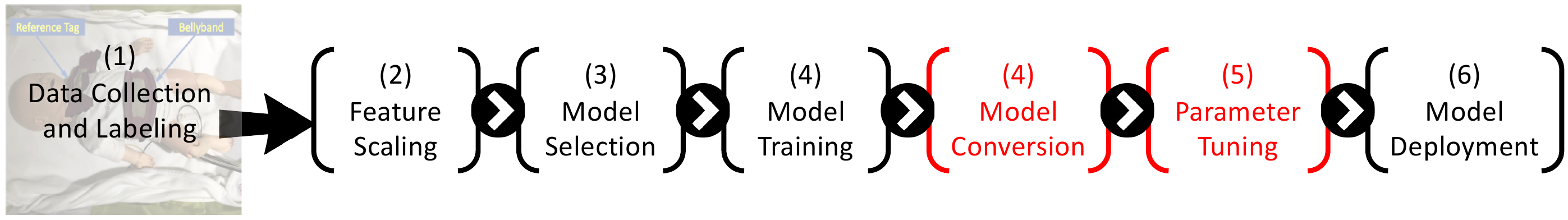}
		%\vspace{-20pt}
	%\end{center}
	\caption{Seven stage pipeline, including the two new stages to process and optimize the SNN model.}% related to this proposal.}
	\label{fig:snn_pipeline}
\end{figure}

To integrate SNN-based respiratory classification into our design pipeline, we introduce two additional stages -- model conversion and SNN parameter tuning, before the SNN model is deployed to perform classification from live data collected from the SimBaby. Figure~\ref{fig:snn_pipeline} shows the new design pipeline.

\subsection{Model Conversion}
In this work, the 1DCNN architecture is converted to SNN in order to execute it on a neuromorphic hardware such as Loihi~\cite{loihi}. The conversion steps are briefly discussed below.
\begin{enumerate}
    \item \emph{ReLU Activation Functions:} This is implemented as the approximate firing rate of a leaky integrate and fire (LIF) neuron.
    
    \item \emph{Bias:} A bias is represented as a constant input current to a neuron, the value of which is proportional to the bias of the neuron in the corresponding analog model.
    
    \item \emph{Weight Normalization:} This is achieved by setting a factor ${\lambda}$ to control the firing rate of spiking neurons.
    
    \item \emph{Softmax:} To implement softmax, an external Poisson spike generator is used to generate spikes proportional to the weighted sum accumulated at each neuron.
    
    \item \emph{Max and Average Pooling:} To implement max pooling, the neuron which fires first is considered to be the winning neuron, and therefore, its responses are forwarded to the next layer, suppressing the responses from other neurons in the pooling function. To implement average pooling, the average firing rate (obtained from total spike count) of the pooling neurons are forwarded to the next layer of the SNN.
    \item \emph{1-D Convolution:} The 1-D convolution is implemented to extract patterns from inputs in a single spacial dimension. A 1xn filter, called a kernel, slides over the input while computing the element-wise dot-product between the input and the kernel at each step.
    \item \emph{Residual Connections:} Residual connections are implemented to convert the residual block used in CNN models such as ResNet. Typically, the residual connection connects the input of the residual block directly to the output neurons of the block, with a synaptic weight of `1'. This allows for the input to be directly propagated to the output of the residual block while skipping the operations performed within the block.
    \item \emph{Flattening:} The flatten operation converts the 2-D output of the final pooling operation into a 1-D array. This allows for the output of the pooling operation to be fed as individual features into the decision making fully connected layers of the CNN model.
    \item \emph{Concatenation:} The concatenation operation, also known as a merging operation, is used as a channel-wise integration of the features extracted from 2 or more layers into a single output.
\end{enumerate}

\begin{figure}[h!]
	\centering
	\vspace{-5pt}
	\centerline{\includegraphics[width=0.99\columnwidth]{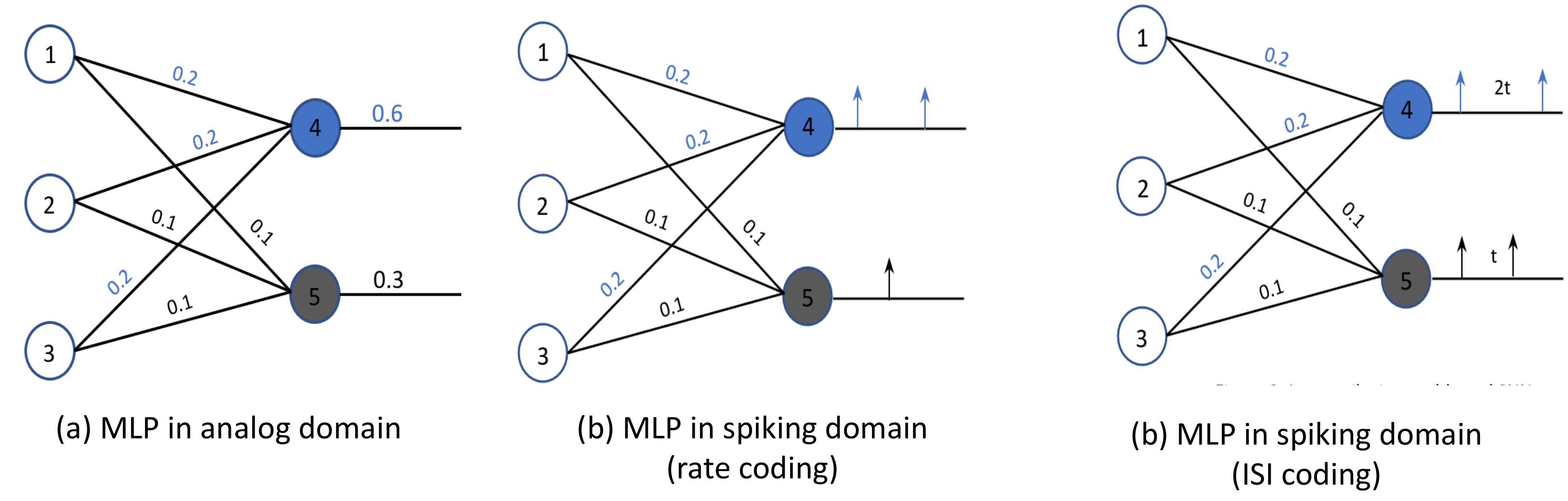}}
	\vspace{-10pt}
	\caption{Example of converting an analog MLP to its spiking equivalent.}
	\vspace{-10pt}
	\label{fig:mlp_example}
\end{figure}

 We now briefly elaborate how an analog operation such as Rectified Linear Unit (ReLU) is implemented using SNN. The output \ineq{Y} of a ReLU activation function is given by
\begin{equation}
    \label{eq:relu_fn}
    \footnotesize Y = \max{0,\sum_i w_i*x_i},
\end{equation}
where \ineq{w_i} is the weight and \ineq{x_i} is the activation on the \ineq{i^\text{th}} synapse of the neuron. To map the ReLU activation function, we consider a particular type of spiking neuron model known as an Integrate and Fire (IF) neuron model. The IF spiking neuron's transfer function can be represented as
\begin{equation}
    \label{eq:if_neuron}
    \footnotesize v_m(t+1) = v_m(t) + \sum_i w_i*x_i(t),
\end{equation}
where \ineq{v_m(t)} is the membrane potential of the IF neuron at time \ineq{t}, \ineq{w_i} is the weight, and \ineq{x_i(t)} is the activation on the \ineq{i^\text{th}} synapse of the neuron at time \ineq{t}. The IF spiking neuron integrates incoming spikes (\ineq{X_i}) and generates an output spike (\ineq{Y_\text{spike}}) when the membrane potential (\ineq{v_m}) exceeds the threshold voltage (\ineq{v_\text{th}}) of the IF neuron. Therefore, by ensuring that the output spiking rate \ineq{Y_\text{spike}} is proportional to the ReLU activation \ineq{Y}, i.e., \ineq{Y_\text{spike}\propto Y}, we accurately convert the ReLU activation to the spike-based model.
To further illustrate this, we consider the multi-layer perceptron (MLP) of Figure~\ref{fig:mlp_example}a and its SNN conversion using rate-based encoding (Figure~\ref{fig:mlp_example}b) and inter-spike interval (ISI) encoding (Figure~\ref{fig:mlp_example}c).

\begin{figure*}[h!]
    \centering
	%\begin{center}
	    %\vspace{-40pt}
		\includegraphics[width=1.99\columnwidth]{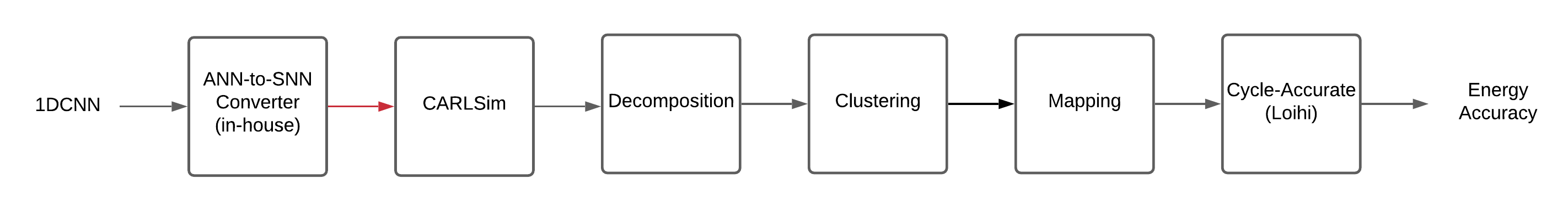}
		%\vspace{-20pt}
	%\end{center}
	\caption{The NeuroXplorer framework~\cite{neuroxplorer}.}% related to this proposal.}
	\label{fig:neuroxplorer}
\end{figure*}

In Figure~\ref{fig:mlp_example}a, neurons 1,2 and 3 are the input neurons and neurons 4 and 5 are the output neurons. To keep the model simple, let us consider the case where the activations of the input neurons 1,2 and 3 are equal to 1. Using Equation~\ref{eq:relu_fn}, we know that the output of neurons 4 and 5 are 0.6 and 0.3, respectively. 
Figures~\ref{fig:mlp_example}b and~\ref{fig:mlp_example}c show the mapped SNN model, using rate-based and inter-spike interval encoding schemes, respectively. In the rate-based model in Figure~\ref{fig:mlp_example}b, the rate of spikes generated is expected to be proportional to the output of neurons 4 and 5 in the MLP. In the case of the ISI-based SNN model, the inter-spike interval of the spikes generated by neurons 4 and 5 is expected to be proportional to the output generated in the MLP, as shown in Figure~\ref{fig:mlp_example}c.

\subsection{SNN Mapping to Neuromorphic Hardware}
The SNN model generated using the conversion approach is analyzed in CARLsim~\cite{carlsim} to generate the following information.
\begin{itemize}
    \item \emph{\textcolor{black}{Spike Data:}} the exact spike times of all neurons in the SNN model.
    \item \emph{\textcolor{black}{Weight Data:}} the synaptic strength of all synapses in the SNN model.
\end{itemize}
The spike and weight data of a trained SNN form the {SNN workload}, which is used in the NeuroXplorer framework~\cite{neuroxplorer} to estimate the energy consumption. Figure~\ref{fig:neuroxplorer} shows the NeuroXplorer framework.

The framework inputs the 1DCNN model and estimates the accuracy and energy consumption of the model on a neuromorphic hardware. Internally, NeuroXplorer first converts the 1DCNN to SNN using the steps outlined in the before. It then simulates the SNN using CARLsim. The extracted workload is first decomposed using the decomposition approach presented in~\cite{esl20}. This is to ensure that the workload can fit on to the resource-constraint hardware. 

Typically, neuromorphic hardware are designed as tile-based architectures~\cite{catthoor2018very}, where each tile can accommodate only a limited number of neurons and synapses. The tiles are interconnected using a shared interconnect such as Network-On-Chip (NoC)~\cite{liu2018neu} or Segmented Bus~\cite{balaji2019exploration}. Therefore, to map an SNN into a tile-based neuromorphic hardware, the model is first partitioned into clusters, where each cluster consists of a proportion of the neurons and synapses of the original machine learning model~\cite{psopart}. Each cluster can then fit onto a tile of the hardware. Then the clusters are mapped to the tiles to optimize one or more hardware metrics such as energy~\cite{twisha_energy,spinemap}, latency~\cite{dfsynthesizer,das2018dataflow,balaji2019ISVLSIframework,balaji2020run,adarsha_igsc}, circuit aging~\cite{twisha_thermal,reneu,ncrtm,song2020case,vts_das,shihao_igsc}, and endurance~\cite{espine,twisha_endurance,song2021improving}. We use the energy-aware mapping technique of~\cite{spinemap}.

Once the clusters of the converted 1DCNN model is placed to the resources of the neuromorphic hardware, we perform cycle-accurate simulations using NeuroXplorer, configured to simulate the Loihi neuromorphic system. Table~\ref{tab:hw_parameters} shows the hardware parameters that are configured in NeuroXplorer.

\begin{table}[h!]
    \caption{Major simulation parameters extracted from Loihi~\cite{loihi}.}
	\label{tab:hw_parameters}
	\vspace{-5pt}
	\centering
	{\fontsize{10}{12}\selectfont
		\begin{tabular}{lp{5cm}}
			\hline
			Neuron technology & 16nm CMOS (original design is at 14nm FinFET)\\
			\hline
			Synapse technology & {HfO${}_2$-based OxRRAM}~\cite{mallik2017design}\\
			\hline
			Supply voltage & 1.0V\\
			\hline
			Energy per spike & 23.6pJ at 30Hz spike frequency\\
			\hline
			Energy per routing & 3pJ\\
			\hline
			Switch bandwidth & 3.44 G. Events/s\\
			\hline
% 			NVM related & 1T-1R\\
% 			& PCM cell SET: 24 cycles\\
% 			& PCM cell RESET: 18 cycles\\
% 			& Program \& Verify: 35 cycles\\
% 			\hline
	\end{tabular}}
\end{table}

\subsection{SNN Parameter Tuning}\label{sec:parameter_tuning}
Unlike the baseline 1DCNN architecture, where model hyperparameters are explored only during model training, SNNs allow parameter tuning on the trained (and converted) model, such that the energy and accuracy space could be explored to generate a solution that satisfies the given energy and accuracy constraints of the target wearable platform. To explore such exploration capabilities, we analyze the dynamics of SNNs.

The membrane potential of a neuron at time \ineq{t} can be expressed as~\cite{maass1997networks}
\begin{equation}
    \label{eq:membrane_potential}
    u(t) = u_0 + a\int_0^t D(s)\cdot w\cdot \sigma(t-s)ds,
\end{equation}
where \ineq{u_0} is the initial membrane potential, \ineq{a} is a positive constant, \ineq{D(s)} is a linear filter, \ineq{w} is the synaptic weight and \ineq{\sigma} represents a sequence of \ineq{N} input spikes, which can be expressed using the Dirac delta function as 
\begin{equation}
    \label{eq:spike_train}
    \sigma(t) = \sum_{i=1}^{N}\delta(t-t_i)
\end{equation}
The membrane potential of a neuron increases upon the arrival of an input spike. Subsequently, the membrane potential starts to decay during the inter-spike interval (ISI). When the neuron is subjected to an input spike train, the membrane voltage keeps rising, building on the undissipated component. When the membrane potential crosses a threshold (\ineq{V_{th}}), the neuron emits a spike, which then propagates to other neurons of the SNN. The spike rate of a neuron can be controlled using this threshold. If the threshold is set too high, less spikes will be generated, meaning that the energy will be lower but so is the accuracy because spikes encode information in SNNs. Therefore, by adjusting the threshold, the design space of accuracy and energy can be explored (see Section~\ref{sec:snn_results}). 

\section{Results and Discussions}
\label{sec:results}
All simulations are performed on a workstation, which has AMD Threadripper 3960X with 24 cores, 128 MB cache, 128 GB RAM, and 2 RTX3090 GPUs. Keras~\cite{keras} is used to implement the baseline 1DCNN, which uses TensorFlow backend~\cite{tensorflow}. QKeras~\cite{coelho2020ultra} is used for training and testing the quantized neural network. Finally, CARLsim~\cite{carlsim} is used for SNN function simulations.

We present our respiratory classification results organized into 1) results for the baseline 1DCNN model (Section~\ref{sec:baseline_results}), 2) results using quantization (Section~\ref{sec:quantization_results}) , and 3) SNN-specific results (Section~\ref{sec:snn_results}).

\subsection{Baseline 1DCNN Performance}\label{sec:baseline_results}
In this section we evaluate the performance of the proposed 1DCNN specified using the following metrics.
% Performance is 
\begin{itemize}
    \item \textbf{Top-1 Accuracy:} This is the conventional accuracy and it measure the proportion of test examples for which the predicted label (i.e., respiratory state) matches the expected label. To formulate top-1 accuracy,
    we introduce the following definitions.
    \begin{itemize}
        \item \textbf{True Positives (TP):} For binary classification problems, i.e., ones with a yes/no outcome (such as the case of respiratory classification), this is the total number of test examples for which the value of actual class is yes and the value of predicted class is also yes. 
        \item \textbf{True Negatives (TN):} This is the total number of test examples for which the value of actual class is no and value of predicted class is also no.
        \item \textbf{False Positives (FP):} This is the total number of test examples for which the value of actual class is no but the value of predicted class is yes.
        \item \textbf{False Negatives (FN):} This is the total number of test examples for which the value of actual class is yes but the value of predicted class is no.
    \end{itemize}
    \begin{equation}
        \label{eq:top_1_accuracy}
        \text{Top-1 Accuracy} = \frac{TP+TN}{TP+FP+FN+TN}
    \end{equation}
    \item \textbf{F1 Score:} To formulate F1 score, we introduce the following definitions.
    \begin{itemize}
        \item \textbf{Precision:} This is the ratio of correctly predicted positive observations to the total predicted positive observations, i.e.,
        \begin{equation}
            \label{eq:precision}
            \text{Precision} = \frac{TP}{TP+FP}
        \end{equation}
        \item \textbf{Recall:} This is the ratio of correctly predicted positive observations to the all observations in actual class, i.e.,
        \begin{equation}
            \label{eq:recall}
            \text{Recall} = \frac{TP}{TP+FN}
        \end{equation}
    \end{itemize}
    F1 Score conveys the balance between the precision and the recall. It is calculated as the weighted average of precision and recall, i.e.,
    \begin{equation}
        \label{eq:f1_score}
        \text{F1 Score} = \frac{2*(\text{Recall} * \text{Precision})}{(\text{Recall} + \text{Precision})}
    \end{equation}
    \item \textbf{AUC:} In machine learning, a receiver operating characteristic (ROC) curve is a graphical plot that illustrates the diagnostic ability of a binary classifier as its discrimination threshold is varied. The area under curve (AUC) measures the two-dimensional area underneath the ROC curve. AUC tells how much the model is capable of distinguishing between classes. Higher the AUC, the better the model is at predicting yes classes as yes and no classes as no.
    \item \textbf{Sensitivity:} This is the true positive rate, i.e., how often the model correctly generates a yes out of all the examples for which the value of actual class is yes. Sensitivity is formulated as
    \begin{equation}
        \label{eq:sensitivity}
        \text{Sensitivity} = \frac{TP}{TP + FN}
    \end{equation}
    \item \textbf{Specificity:} This is the true negative rate, i.e., how often the model correctly generates a no out of all the examples for which the value of actual class is no. Specificity is formulated as
    \begin{equation}
        \label{eq:sensitivity}
        \text{Specificity} = \frac{TN}{TN + FP}
    \end{equation}
\end{itemize}

Table~\ref{tab:sota} compares the classification performance using the proposed 1DCNN against three state-of-the-art approaches -- 1) Support Vector Machine (SVM) classifier of~\cite{mongan2015statistical}, 2) Logistic Regression (LR) classifier of~\cite{acharya2018ensemble}, and 3) Random Forest classifier of~\cite{acharya2018ensemble}. We make the following four key observations.

\begin{table}[h!]
	\renewcommand{\arraystretch}{1.2}
	\setlength{\tabcolsep}{2pt}
	\caption{Comparison with State-of-the-Art Approaches.}
	\label{tab:sota}
	%\vspace{-10pt}
	\centering
	\begin{threeparttable}
	{\fontsize{7}{10}\selectfont
	    %\vspace{-10pt}
		\begin{tabular}{l|ccccc}
			\hline
			\textbf{Classification Technique} & \textbf{Top-1 Accuracy} & \textbf{F1 Score} & \textbf{AUC} & \textbf{Sensitivity} & \textbf{Specificity}\\
			\hline
			SVM & 92.34\% & 0.91 & 0.92 & 0.93 & 0.92\\
			LR & 91.60\% & 0.91 & 0.90 & 0.90 & 0.92\\
			RF & 93.40\% & 0.92 & 0.90 & 0.92 & 0.93\\
			\hline
			1DCNN (proposed) & 97.15\% & 0.98 & 0.98 & 0.96 & 0.99\\
			\hline
	\end{tabular}}
	\end{threeparttable}
	%\vspace{-10pt}
\end{table}

First, the proposed 1DCNN has the highest top-1 accuracy of all the evaluated techniques (higher top-1 accuracy is better). The top-1 accuracy of 1DCNN is better than SVM by 5.2\%, LR by 6.0\%, and RF by 4.0\%.

Second, the proposed 1DCNN has the highest F1 score of all the evaluated techniques (higher F1 score is better). The F1 score of 1DCNN is higher than SVM by 7.7\%, LR by 7.7\%, and RF by 6.5\%.

Third, the proposed 1DCNN has the highest AUC of all the evaluated techniques (higher AUC score is better). The AUC score of 1DCNN is higher than SVM by 6.2\%, LR by 8.5\%, and RF by 8.5\%.

Fourth, the proposed 1DCNN has the highest sensitivity of all the evaluated techniques (higher sensitivity score is better). The sensitivity score of 1DCNN is higher than SVM by 3.2\%, LR by 6.7\%, and RF by 4.3\%.

Finally, the proposed 1DCNN has the highest specificity of all the evaluated techniques (higher specificity score is better). The specificity score of 1DCNN is higher than SVM by 7.6\%, LR by 7.6\%, and RF by 6.4\%.

%We observe that the proposed 1DCNN model outperforms all the existing approaches to respiratory classification. On average, the proposed model has x\% higher accuracy, y\% higher F1 score, z\% higher AUC, w\% higher sensitivity, and u\% higher specificity. 
The reason for high performance using the proposed 1DCNN model is two-fold. First, we perform intelligent feature selection from the data collected using sensors on the SimBaby programmable infant mannequin. Second, we perform hyperparameter optimization with neural architecture search to generate a model that gives the highest classification accuracy using the selected hyperparameters.

\begin{figure}[h!]
    \centering
	%\begin{center}
	    %\vspace{-40pt}
		\includegraphics[width=0.99\columnwidth]{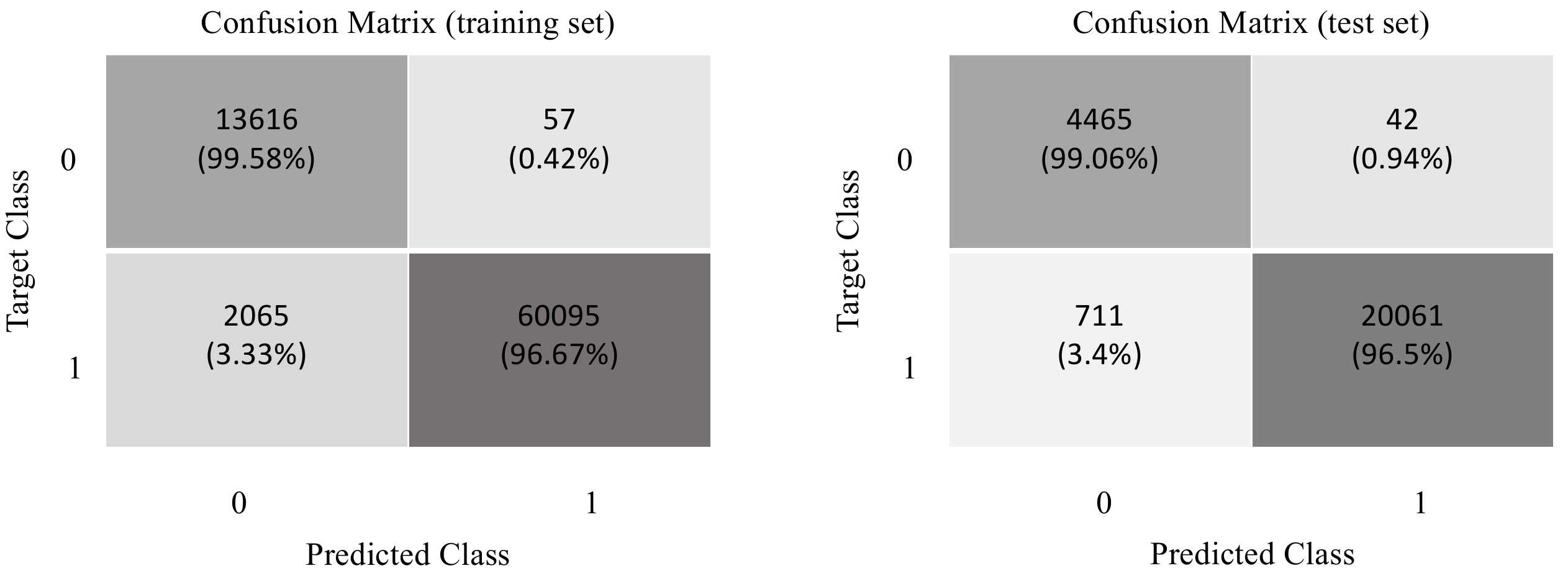}
		%\vspace{-20pt}
	%\end{center}
	\caption{Confusion matrix for the 1DCNN model.}% related to this proposal.}
	\label{fig:confusion_matrix}
\end{figure}

To give further insight to the improvement, Figure~\ref{fig:confusion_matrix} shows the confusion matrix obtained for the training and test sets. We observe that the proposed 1DCNN model has very low false positives and false negatives, which are critical for respiratory classification in premature newborn infants.

\subsection{Quantization Results}\label{sec:quantization_results}

Table~\ref{tab:qkeras} reports the top-1 accuracy (\%), energy (in pJ), and model size (in bits) with 2-bit, 4-bit, 8-bit, 16-bit, and 32-bit precision for the model parameters. For comparison, we have included results using the baseline 1DCNN, which uses full 64-bit precision for the model parameters. We make the following three key observations.

\begin{table}[h!]
	\renewcommand{\arraystretch}{1.2}
	\setlength{\tabcolsep}{2pt}
	\caption{Model Quantization Results.}
	\label{tab:qkeras}
	%\vspace{-10pt}
	\centering
	\begin{threeparttable}
	{\fontsize{9}{12}\selectfont
	    %\vspace{-10pt}
		\begin{tabular}{l|ccc}
			\hline
			\textbf{Quantization} & \textbf{Top-1 Accuracy} & \textbf{Energy (pJ)} & \textbf{Model Size (bits)}\\
			\hline
			2-bit/parameter & 88.93\% & 7,089 & 92,258\\
			4-bit/parameter & 88.98\% & 15,994 & 184,516 \\
			8-bit/parameter & 93.00\% & 29,871 & 369,032 \\
			16-bit/parameter & 96.55\% & 57,640 & 738,064\\
			32-bit/parameter & 97.03\% & 113,386 & 1,476,128 \\
			\hline
			Baseline 1DCNN & 97.15\% & 134,613 & 2,952,256\\
			\hline
	\end{tabular}}
	\end{threeparttable}
	%\vspace{-10pt}
\end{table}

First, top-1 accuracy reduces with a reduction in the bit precision (higher accuracy is better for respiratory classification in premature newborn infant). With 2-bit, 4-bit, 8-bit, 16-bit, and 32-bit precision, the top-1 accuracy is lower than the baseline 64-bit precision by 8.5\%, 8.4\%, 4.3\%, 0.6\%, and 0.1\%, respectively. The top-1 accuracy with 32-bit precision is comparable to 64-bit precision. 

Second, energy reduces with a reduction of the bit  precision (lower energy is better for respiratory classification in wearables due to limited battery as we motivated in Section~\ref{sec:intro}).
With 2-bit, 4-bit, 8-bit, 16-bit, and 32-bit precision, energy is lower than the baseline 64-bit precision by 94.7\%, 88.1\%, 77.8\%, 57.2\%, and 15.8\%, respectively. These results show the significant reduction in energy achieved using quantization. Lower energy leads to longer battery life in wearables.
Finally, model size also reduces with a reduction in the bit precision (lower model size is better for wearables due to their limited storage availability). 
With 2-bit, 4-bit, 8-bit, 16-bit, and 32-bit precision, energy is lower than the baseline 64-bit precision by 96.8\%, 93.7\%, 87.5\%, 75.0\%, and 50.0\%, respectively.

%To further analyze the energy bottleneck, Figure~\ref{fig:energy_bottleneck} shows the energy bottleneck of different layers in the 1DCNN for the 5 different settings of the bit precision compared to the baseline 1DCNN model.

We conclude that to reduce energy and model size for respiratory classification in wearables, quantization techniques can lead to a significant reduction in accuracy.

\subsection{SNN-Related Results}\label{sec:snn_results}

Table~\ref{tab:snn_results} reports the top-1 accuracy and energy results using the proposed SNN-based approach compared to the baseline 1DCNN, 2-bit, and 8-bit quantized model. We make the following three key observations.

\begin{table}[h!]
	\renewcommand{\arraystretch}{1.2}
	\setlength{\tabcolsep}{2pt}
	\caption{SNN Accuracy and Energy Results.}
	\label{tab:snn_results}
	%\vspace{-10pt}
	\centering
	\begin{threeparttable}
	{\fontsize{10}{14}\selectfont
	    %\vspace{-10pt}
		\begin{tabular}{l|cc}
			\hline
			\textbf{Model} & \textbf{Top-1 Accuracy} & \textbf{Energy (pJ)}\\
			\hline
			Baseline 1DCNN & 97.15\% & 134,613\\
			2-bit Quantized 1DCNN & 88.93\% & 7,089\\
			8-bit Quantized 1DCNN & 93.00\% & 29,871\\
			\hline
			SNN & 93.33\% & 7,282 \\
			\hline
	\end{tabular}}
	\end{threeparttable}
	%\vspace{-10pt}
\end{table}

First, the top-1 accuracy of the proposed SNN is only 4\% lower than the baseline 1DCNN model with 18x lower energy. 
Second, compared to 2-bit quantized model, the top-1 accuracy is 5\% higher, while the energy is only 2\% higher. Finally, compared to 8-bit quantized model, the top-1 accuracy is comparable while the energy is 4x lower. We conclude that SNN-based respiratory classification achieves the best tradeoff in terms of top-1 accuracy and energy. To achieve similar accuracy, SNN can lead to 4x reduction in energy, which is a critical consideration for respiratory classification on wearables. The following results are reported to give further insight into these improvements.

\subsubsection{SNN Accuracy Compared to 1DCNN}
\begin{figure}[h!]
    \centering
	%\begin{center}
	    %\vspace{-40pt}
		\includegraphics[width=0.99\columnwidth]{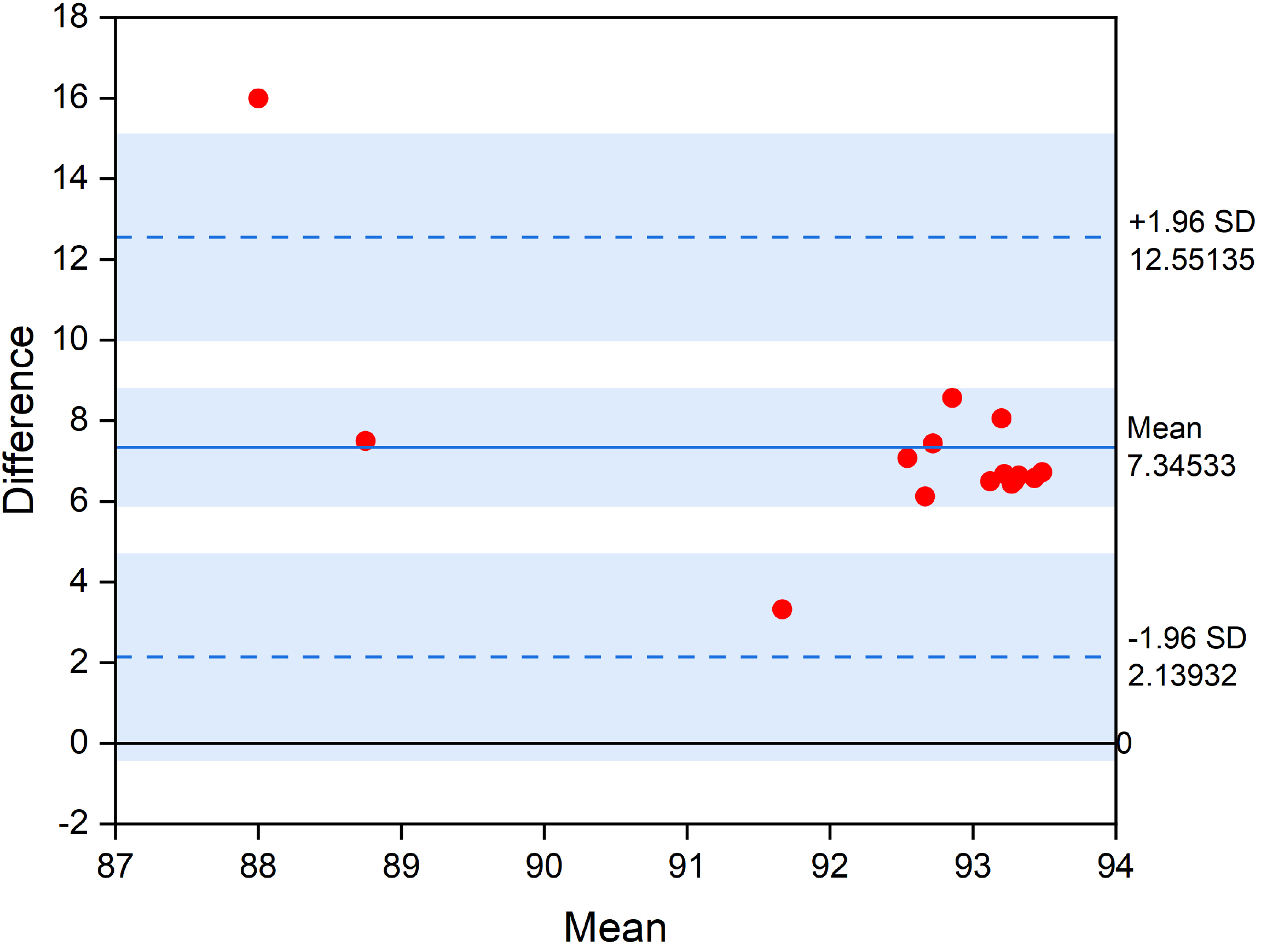}
		%\vspace{-20pt}
	%\end{center}
	\caption{Bland-Altman plot comparing the accuracy of different SNN solutions against the baseline 1DCNN model.}% related to this proposal.}
	\label{fig:bland_altman}
\end{figure}

Figure~\ref{fig:bland_altman} shows the Bland-Altman plot comparing the accuracy of SNN solution against the baseline 1DCNN model. Bland–Altman plots are extensively used to evaluate the agreement among two models, each of which produced some error in their predictions. As can be seen from the plot, the average accuracy difference between the 1DCNN and the converted SNN is 7.3\%, while the minimum and maximum accuracy difference are 2.1\% and 12.5\%, respectively.

\subsubsection{Design Space Exploration with SNN Parameters} We perform design-space explorations to identify SNN model parameters that give the best tradeoff in terms of energy and accuracy.

\begin{figure}[h!]
    \centering
	%\begin{center}
	    %\vspace{-40pt}
		\includegraphics[width=0.99\columnwidth]{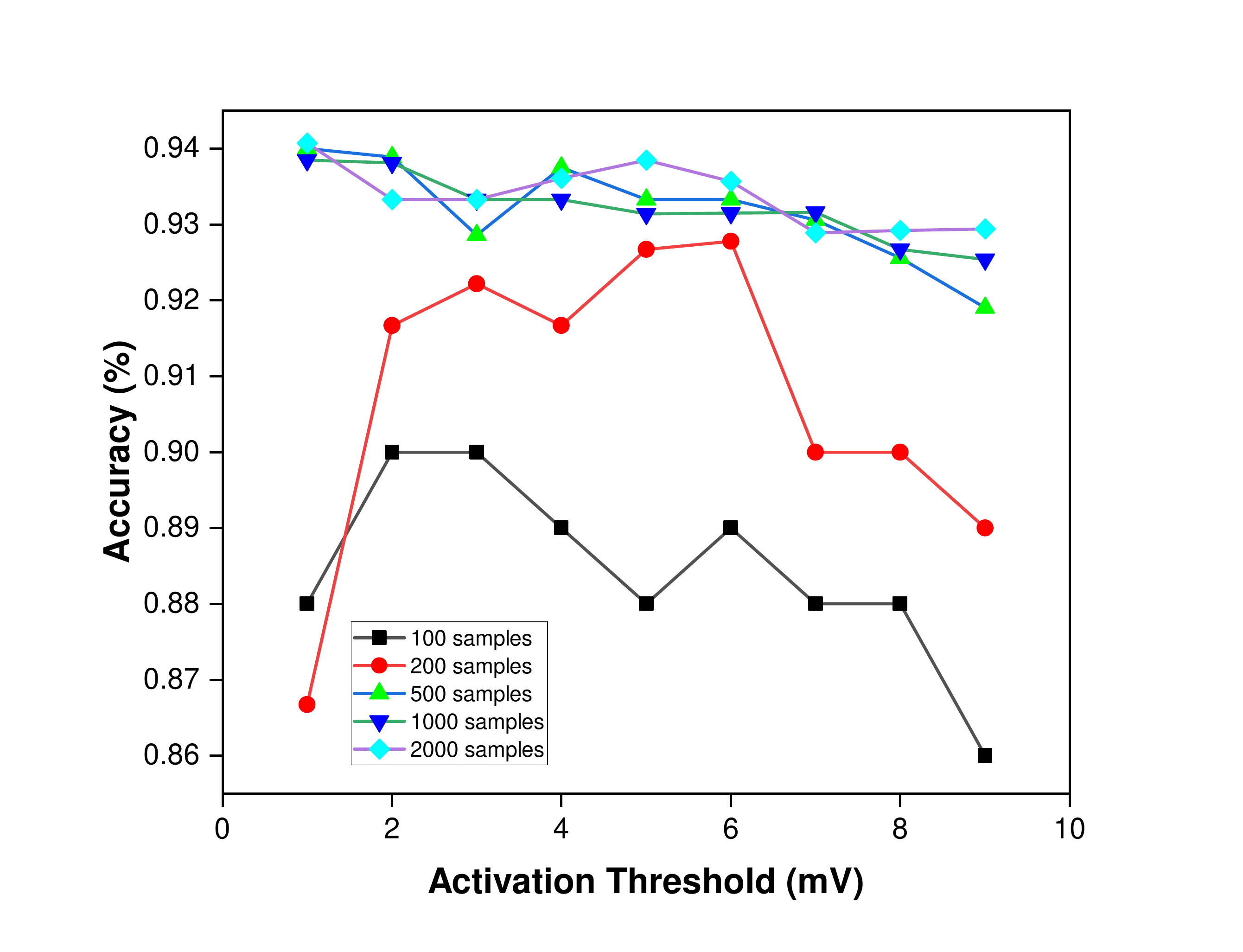}
		%\vspace{-20pt}
	%\end{center}
	\caption{Accuracy impact with varying the activation.}% related to this proposal.}
	\label{fig:activation}
\end{figure}

In spiking neural networks, a spike is not fired by a neuron unless the specified activation threshold voltage is attained. This implies, the larger the firing threshold voltage, the more selectively a neuron is fired while communicating between each layer. To demonstrate this,
Figure~\ref{fig:activation} shows the variation in accuracy as a function of the activation threshold (\ineq{V_{th}}) (see Section~\ref{sec:parameter_tuning}). 
%Results are reported for varying number of samples. We make the following three key observations.

In the figure, we vary the activation threshold, across different number of test samples. We observe that for smaller number of test samples, as we increase the threshold voltage for firing spikes, the performance accuracy varies highly. But for larger size of test samples, the performance accuracy does not vary as much, when we increase the threshold voltage. 

For larger test sample sizes (500, 1000, 2000 etc.) the performance accuracy varies between 92\% and 94\%. When we use small sample sizes (e.g., 100 and 200) the prediction accuracy varies between 86\% and 92\%. We observe that the firing threshold is most ideal when set within 1-3mV for smaller test samples and within 1-5mV for larger test samples.  

%\textcolor{red}{Add the observations here.}

% \section{Discussion}
% \label{sec:results}
% \input{sections/results.tex}

\section{Conclusions}
\label{sec:conclusion}
We propose a Deep Learning enabled wearable monitoring system for premature newborn infants, where {respiratory cessation is predicted} using signals that are collected wirelessly from a non-invasive wearable Bellypatch put on the infant’s body. To this end, we developed an end-to-end design pipeline involving five stages  -- data collection, feature scaling, model selection, training, and deployment. The deep learning model is a 1D convolutional neural network (1DCNN), the parameters of which are tuned using a grid search methodology. Our design achieved 97.15\% accuracy compared to state-of-the-art statistical approaches. To address the limited energy in wearable settings, we evaluate model compression techniques such as quantization. We show that such techniques can lead to a significant reduction in respiratory classification accuracy in order to minimize energy. To address this important problem, we propose a novel Spiking Neural Network (SNN)-based respiratory classification technique, whicch can be implemented efficiently on an event-driven neuromorphic hardware. SNN-based respiratory classification involves two additional pipeline stages -- model conversion and parameter tuning. Using pulseoxiometer data collected from a  Laerdal SimBaby programmable infant mannequin, we demonstrate 93.33\% respiratory classification accuracy with 18x lower energy compared to the conventional 1DCNN model. 

We \textbf{conclude} that SNNs have the potential to implement respiratory classification and other machine learning tasks on energy-constrained environments such as wearable systems.\\

Our research results are based upon work supported by the National Science Foundation Division of Computer and Network Systems under award number CNS-1816387.  Any opinions, findings, and conclusions or recommendations expressed in this material are those of the author(s) and do not necessarily reflect the views of the National Science Foundation. Research reported in this publication was supported by the National Institutes of Health under award number R01 EB029364-01. The content is solely the responsibility of the authors and does not necessarily represent the official views of the National Institutes of Health.

% \section*{Future Work}
% This section will be filled in after the double blind review process.

%\section*{References}
\bibliographystyle{IEEEtran}
\IEEEtriggeratref{41}
\bibliography{commands,disco,internal,external,mongan}

\end{document}